\documentstyle{article}
\title{Erratum\\
Is it possible to infer the equation of state of a mixture of 
hard discs from that of the one-component system?}

\author{ANDR\'ES SANTOS\\
Departamento de F\'{\i}sica, Universidad  de  Extremadura, \\
E-06071 Badajoz, Spain
}
\date{({\em Molec. Phys.}, 1999, {\bf 96}, 1185--1188)}
\begin{document}
\maketitle

The numerical values in the sixth and seventh columns of table 1 of the paper 
are not correct. Consequently, the comments made in the paper about the better 
performance of the equation of state (12) proposed in  [4] over the equation of 
state (6) proposed in the paper are wrong. The corrected version of table 1 is 
reprinted here.
In view of this table, it is quite apparent that the results obtained from 
equation (6) are practically indistinguishable from those obtained from equation 
(12). For the cases considered in the table, both equations of state differ by 
less than 0.05\% when combined with equation (14) for the one-component system 
and by less than 0.02\% when combined with equation (15) for the one-component 
system. These differences increase, but are kept relatively small, when  
mixtures more asymmetric are considered. For instance, in the case $\alpha=0.1$, 
$x_1=0.01$ and $\eta=0.6$, the values of $p\sigma_1^2/k_BT$ predicted by 
(6)+(14), (6)+(15), (12)+(14) and (12)+(15) are 183.87, 186.26, 185.40 and 
186.53, respectively. Table 1 also shows that, in general, the use of equation 
(15) gives a better agreement with the simulation data typically for 
$\eta<0.55$, while the use of equation (14) is preferable for $\eta>0.55$.
\begin{table}
\begin{tabular}{cllcccccc}
\hline\\
$\alpha$&$x_1$&$\eta$&Simul.&({5})
&({6})+({14})&
({6})+({15})&
({12})+({14})&
({12})+({15})\\
\hline\\
0.9&0.48&0.54&3.70&3.72&3.6532& 3.7057&3.6531&  3.7058\\
0.9&0.49&0.63&6.54&6.78&6.6810& 6.6867&6.6806&  6.6870\\
0.8&0.65&0.55&4.05&4.06&3.9985& 4.0511&3.9967&  4.0515\\
0.8&0.315&0.55&4.72&4.71&4.6322&4.6970& 4.6338&  4.6972\\
0.8&0.52&0.60&5.88&5.96&5.8708&5.9126& 5.8688&  5.9136\\
0.8&0.315&0.60&6.33&6.55&6.4428& 6.4958&6.4461&  6.4954\\
0.7&0.546&0.55&4.55&4.57&4.5007& 4.5584&4.4982&  4.5593\\
0.8&0.351&0.481&3.00&3.03&2.9762& 3.0234&2.9766&  3.0237\\
0.8&0.351&0.532&4.00&4.14&4.0642& 4.1248&4.0650&  4.1251\\
0.8&0.351&0.548&4.50&4.57&4.4956& 4.5586&4.4966&  4.5588\\
0.8&0.351&0.564&5.00&5.07&4.9824&  5.0458&4.9837& 5.0460\\
0.8&0.351&0.579&5.50&5.59&5.4975& 5.5585&5.4991&  5.5586\\
\hline
\end{tabular}
\caption{Comparison of  Monte Carlo simulation values [9] 
of the quantity $p\sigma_1^2/k_BT$ with the predictions of 
several equations of state:  equation ({5}) proposed by 
Wheatley [3],  equation ({6}) derived in this 
paper, complemented by equations ({14}) and ({15}), 
and equation ({12}) proposed in [4],
complemented by equations ({14}) and ({15}).
\label{table1}}
\end{table}

\end{document}